\documentclass[journal,oneside]{IEEEtran}
\usepackage{times,amsmath,mathtools,epsfig,amsfonts,amssymb,graphicx,url,cite,colortbl,bm,upgreek,multirow,booktabs,numprint,subfigure}

\usepackage{standalone}
\usepackage{tikz,pgf}
\usepackage[T1]{fontenc}
\usepackage{pgfplots}
\usepackage{grffile}
\pgfplotsset{compat=newest}
\usetikzlibrary{plotmarks}
\usetikzlibrary{arrows.meta}
\usepgfplotslibrary{patchplots}
\pgfplotsset{compat=newest}
\newlength\fwidth
\begin{document}
	
	\hyphenation{}
	\title{Electromagnetic-Thermal Analyses of Distributed Antennas Embedded into a Load Bearing Wall}
\author{Lauri V\"{a}h\"{a}-Savo,~\IEEEmembership{Student~Member,~IEEE,}
        Katsuyuki Haneda,~\IEEEmembership{Member,~IEEE,} 
        Clemens Icheln,~
	Xiaoshu L\"{u}~
	
\thanks{L.  V\"{a}h\"{a}-Savo, K. Haneda and C. Icheln are with the Department of Electronics and Nanoengineering, Aalto University-School of Electrical Engineering, Espoo FI-00076, Finland (e-mail: lauri.vaha-savo@aalto.fi).}
\thanks{X. L\"{u} is with University of Vaasa, Finland.}
\thanks{The work has been supported by the Academy of Finland Research Project ``Signal-Transmissive-Walls with Embedded Passive Antennas for Radio-Connected Low-Energy Urban Buildings (STARCLUB)," decision \#323896.}

}
	
	\maketitle
	\begin{abstract}
The importance of indoor mobile connectivity has increased during the last years, especially during the Covid-19 pandemic. In contrast, new energy-efficient buildings contain structures like low-emissive windows and multi-layered thermal insulations which all block radio signals effectively. To solve this problem with indoor connectivity, we study passive antenna systems embedded in walls of low-energy buildings. We provide analytical models of a load bearing wall along with numerical and empirical evaluations of wideband back-to-back antenna spiral antenna system in terms of electromagnetic- and thermal insulation. The antenna systems are optimized to operate well when embedded into load bearing walls. Unit cell models of the antenna embedded load bearing wall, which are called {\it signal-transmissive walls} in this paper, are developed to analyze their electromagnetic and thermal insulation properties. We show that our signal-transmissive wall improves the electromagnetic transmission compared to a raw load bearing wall over a wide bandwidth of 2.6-8 GHz, covering most of the cellular new radio frequency range 1, without compromising the thermal insulation capability of the wall demanded by the building regulation. Optimized antenna deployment is shown with $22$~dB improvement in electromagnetic transmission through the load bearing wall.
	\end{abstract}
	
	
\section{Introduction} 
\label{sec:introduction}

The fifth-generation cellular networks, which have been rolled out during the last couple of years, promise higher data rates compared to legacy systems. With new radio (NR) frequency range 1 (FR1), where the frequencies are below $6$~GHz, the capacity can be increased up to 20 times higher than the fourth generation systems~\cite{Nokia5G_18}. At the same time providing sufficient indoor coverage has become more challenging \cite{Zinwave17}. During the Covid-19 pandemic, the use of mobile internet has risen drastically all over Europe~\cite{wood_20} and in the world. The problems with indoor coverage become increasingly serious in energy-efficient buildings. They have e.g. multilayered thermal insulating layers, low-emission airtight windows, and other signal barriers which all make the outdoor-to-indoor (O2I) communication challenging, if not impossible~\cite{Rodriguez14_VTCFall, Haneda16_VTCSpring}. This trend will be strengthened as international building regulations like~\cite{EUDirective} have been taken into national regulation. The goal of the regulation is to move towards zero-energy buildings which makes the cellular coverage by outdoor base stations even more challenging.

To improve indoor coverage, both active and passive solutions have been studied in recent years. Active solutions include indoor base stations~\cite{Yunas15, Zakaria19_CMT} and repeaters \cite{Haneda10_EuCAP, Rigelsford14_EuCAP, Ntontin19_SyS}. The problem with the active solutions is that those often need a radio-frequency-over-fiber network. These are expensive systems and need additional energy to operate. The passive solutions are often e.g. low-emissivity windows with frequency-selective surfaces~\cite{Asp15_EW} or signal slots~\cite{SignalWindow, Lilja19_GPD}. The passive solution does not need extra energy to operate, but the problem is often that those have relatively narrow passbands in the radio frequency (RF). In addition at least to our best knowledge, there is no commonly used passive solution that could operate over the whole lifespan of the building.   

In this paper, we improve the work of our own~\cite{vahasavo_EuMC21} and present a numerical and empirical evaluation of passive antenna systems embedded into a building wall called the {\it signal-transmissive wall}. The main purpose of the embedded antenna system is to create a pathway for the RF signals to propagate into a low-energy building with lower penetration losses. This concept was introduced with an antenna system at a point frequency in~\cite{vahasavo_EuMC21}, but in this paper, we introduce a wideband spiral antenna system that significantly reduces the penetration loss compared to load bearing wall. The antenna system is also designed to keep the thermal insulation of the wall to the level stated by the nation-building code of Finland~\cite{buildingcode}. In this paper, we show first time the optimized deployment of an ordinal antenna system on a wall through multi-physics criterion.

In summary, the novel contributions of the present paper are summarized as the following three-fold.
\begin{enumerate}
 \item We developed a wideband back-to-back spiral antenna system embedded onto the wall, leading to a signal-transmissive wall operating over a large portion of NR FR1 band i.e. between $2.6$ and $8$~GHz;

 \item We perform an analytical, numerical, and empirical evaluation of the electromagnetic- and thermal insulation of the wall to evaluate the efficacy of the signal-transmissive wall; and finally,
 
 \item We perform a numerical study showing the effect of antenna system installation on the reduction of electromagnetic transmission losses and maintained thermal insulation.
\end{enumerate}

The rest of the paper is organized as follows: Section II defines the load bearing wall of modern buildings and analytical expressions of electromagnetic and thermal properties of the wall. Section III introduces the design of a spiral antenna system embedded into the load bearing wall, along with experimental verification of the spiral antenna design in free space. Section IV shows a comparison of electromagnetic and thermal transmission between the bare load bearing wall and the signal-transmissive wall. The electromagnetic properties of the designed signal-transmissive wall is verified experimentally. After that, the optimized installation of the antenna systems are numerically evaluated through an electromagnetic-thermal criterion. Section~V concludes the work and gives some insights for future works.

\section{Electromagnetic and thermal insulation of a load bearing wall}
\label{sec:Electromagnetic_insulation}

	\begin{figure}[ht]
		\begin{center}
			\includegraphics[scale=0.33]{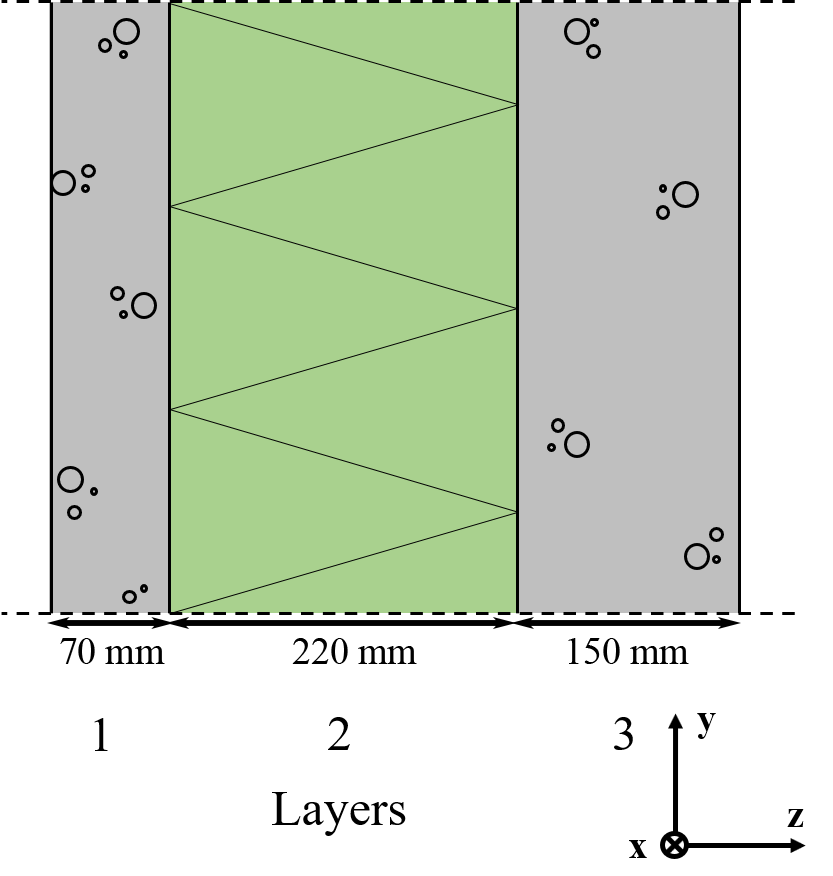}
			\caption{A concrete sandwich panel for load bearing wall. The panel is made of 70\,mm-thick reinforced concrete (layer 1), 220\,mm-thick rock wool for thermal insulation (layer 2), and 150 mm-thick reinforced concrete for structural bearing (layer 3).}\label{fig:wall}
		\end{center}
	\end{figure}

A general structure of a bare load bearing wall that we study as an example in this paper is illustrated in Fig.~\ref{fig:wall}\footnote{A concrete wall can be reinforced by adding a rebar net. Effects of this net are slightly increased penetration losses and cut-off frequency that appears typically below 1 GHz RF. The net does not impact the thermal insulation either, so we do not consider them in this study.}. The wall is made from $220$~mm thick rock wool layer for thermal insulation, which is sandwiched between $70$ and $150$~mm thick concrete layers. The following summarizes analytical models of the electromagnetic and thermal transmission properties of the wall, along with its comparison with numerical simulations to cross-validate the two approaches. This validation shows that our numerical method is in line with the theory and hence has the fidelity needed to perform the analysis in this work. It is important to remember that, the goal of this work is to improve electromagnetic transmission and at the same time maintain thermal insulation so both analyses are crucial for the success of this study.

\subsection{Electromagnetic insulation of the wall}

\subsubsection{Analytical model}
The electrical parameters of concrete and rock wool are defined from 1 to 100~GHz in ITU-R P.2040 \cite{ITU-R_P2040}. The permittivities of different building materials are defined as~\cite{ITU-R_P2040}
\begin{equation}
\label{eq:e}\epsilon_{\rm r}(f)= \epsilon'-j \epsilon'' = a f^b -j \frac{cf^d}{\epsilon_0 \omega},
\end{equation} 
where the $a$, $b$, $c$ and $d$ are summarized in Table \ref{table:param} that determine the frequency dependent permittivity while $f$ is the RF in GHz. We model the wall as a three-layered dielectric structure that is infinitely long in $x$ and $y$ directions as defined by the coordinate system in Fig.~\ref{fig:wall}. 

\subsubsection{Numerical study}

Electromagnetic insulation of the load bearing wall was numerically simulated using the \textit{CST Studio Suite}\footnote{ https://www.3ds.com/products-services/simulia/products/cst-studio-suite/}. An infinitesimally large load bearing wall is considered along with a plane electromagnetic wave incidence. The infinitely large wall can be simulated using unit cell boundary conditions with two Floquet ports~\cite{Tretyakov_03}
which are placed a quarter wavelength distance away from the outdoor and indoor facing sides of the wall which is calculated from the smallest frequency in the simulation. Thereby the small unit cell is copied around itself in $\pm x$ and $\pm y$ directions, making it virtually infinitely large. This means that if an antenna system is placed on a unit cell, it is repeated with the rest of the unit cell and antenna separation will be the same as the $x$ and $y$ dimensions of the unit cell.

\begin{table}[t]
	\begin{small}
	\begin{center}
		\caption{Dimensional and electrical properties of the load bearing wall}
		\label{table:param}
		\begin{tabular}{c|c|c} \hline
			Layer \# /		& Thickness 	& Relative permittivity model		\\ 
			Material		& [mm] 		& $\epsilon_{\rm r}$ in~\eqref{eq:e}~\cite{ITU-R_P2040}  \\ \hline
			\#1	Concrete		& 70			& $a=5.24$, $b=0$,  \\ 
			& & $c=4.6 \times 10^{-2}$, $d=0.7822$ \\ \hline
			\#2	Rock wool		& 220		& $a=1.48$, $b=0$, \\
				& & $c=1.1 \times 10^{-3}$, $d=1.0750$ 	 \\ \hline
			\#3	Concrete		& 150		& $a=5.24$, $b=0$, \\
			& & $c=4.6 \times 10^{-2}$, $d=0.7822$ 	\\ \hline 
		\end{tabular}
	\end{center}
	\end{small}
\end{table}

We assume a right-handed circularly polarized incoming plane wave. Since the unit cell is larger than the wavelength, the higher-order modes must be considered on the receiving side to ensure that all the power coming from the wall is captured. In the simulations, the entire RF band of interests is split into $1$~GHz sub-bands to ensure the feasible simulation time. More modes need to be considered on the receiving side at the higher RF. In this study the number of Floquet modes $N_{\rm F} = 50-100$ is enough to model all the power leaving the wall. The total transmission of the electromagnetic fields through the wall can be calculated by integrating all the mode coefficients in the receiving port as $T = \sqrt{\sum{|T_i|^2}}$, $i = 1 \cdots N_{\rm F}$, given by numerical simulations in the CST Studio.

\begin{figure}[ht]
		\begin{center}
			\includegraphics[scale=0.55]{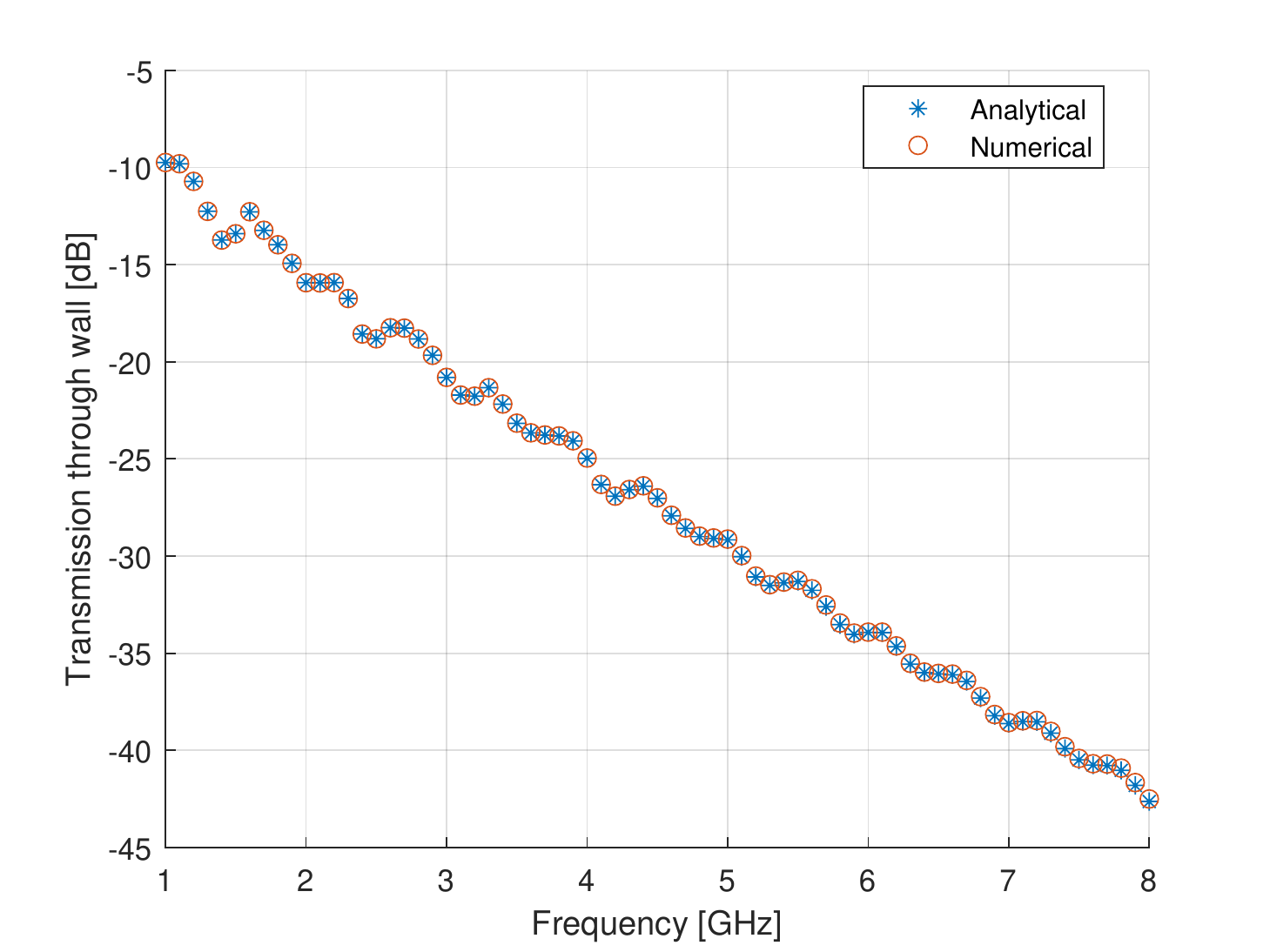}
			\caption{Electromagnetic transmission through a bare load bearing wall, comparison between analytical calculation and numerical simulation.}\label{fig:comparison}
		\end{center}
	\end{figure}

The transmission coefficients of the bare load bearing wall derived from the analytical model and numerical simulations are compared to cross-validate both approaches. As we see from Fig.~\ref{fig:comparison} the electromagnetic transmission coefficients from the two approaches agree perfectly.

\subsection{Thermal insulation of the wall}
\subsubsection{Analytical Study}
Thermal insulation of the wall is often described by conductive heat flux. In the case of steady-state, the heat flux density can be evaluated by using a sample wall, whose dimensions in $x$ and $y$ directions are much larger than in the $z$ direction, so that one-dimensional steady-state heat conduction can be assumed. The heat flux depends on the types of materials and the number of wall layers as well as the temperature difference between different sides of the wall sample. Therefore we use a heat transfer rate per temperature, also called a {\it U-value} and {\it thermal transmittance}, as an evaluation metric. Walls with smaller U-values are preferred for energy-efficient buildings. The thermal transmittance is defined as the inverse of the thermal resistance as 
\begin{equation}
U = \frac{1}{R_{\rm tot}} = \frac{1}{R_{\rm si}+R_{\rm n}+R_{\rm se}},
\end{equation}
where $R_{\rm si}$ and $R_{\rm se}$ are indoor-facing and outdoor-facing thermal surface resistances. The thermal resistance of a wall is affected by radiation from heat sources influencing the wall, e.g., the sun and electrical equipment. Also, air convection affects the thermal surface resistances on both indoor- and outdoor-facing sides of the wall. The ISO6946 standard~\cite{ISO6946} for building components and elements recommends that evaluation of thermal transmittance of walls is made with $R_{\rm si} = 0.13$ and $R_{\rm se} = 0.04$~$\rm m^2\cdot K/W$. Thermal resistance $R_{\rm n}$ of the multi-layered wall can analytically be derived for a simple wall made of layers of $N$ slabs by
\begin{equation}
R_{\rm n} = \sum_{n=1}^N\frac{d_n}{\lambda_n},
\end{equation}
where $d_n$ is the thickness of layer $n$ and $\lambda_n$ is a thermal conductivity $[\rm W/(m\cdot K)]$ of layer $n$.\footnote{A symbol for thermal conductivity $\lambda$ should not be confused with the wavelength of electromagnetic waves in this paper. We follow the notation of ISO6046~\cite{ISO6946} for the symbol $\lambda$.}
\subsubsection{Numerical Study}

To simulate the thermal transmittance of the wall, \textit{Comsol Multiphysics heat transfer module}\footnote{www.comsol.com} was used. The same unit cell model was used as in the electromagnetic simulations. The values of thermal conductivity can be found in Table~\ref{table:param} for each layer of the bare load bearing wall shown in Fig~\ref{fig:wall}. Convective heat fluxes $q_{\rm si}=T_{\rm si}/R_{\rm si}$ and $q_{\rm se}=T_{\rm se}/R_{\rm se}$ as a boundary condition were assigned for indoor and outdoor facing sides of the wall, where $T_{\rm si} = 293$~K and $T_{\rm se} = 271$~K are the indoor and outdoor facing surface temperature, respectively. These convective heat fluxes work as a source of our thermal simulation. For the four sides of the unit cell facing towards $\pm x$- and $\pm y$-directions in Fig.~\ref{fig:wall} thermal insulation boundaries were assigned so that the wall was virtually infinitely large along with those directions. The thermal insulation boundary ensures that the normal component of conductive heat flux on the boundary is zero, i.e., no heat is dissipating on those four sides. To verify the simulation model we compared the simulated thermal transmittance to the analytically calculated one. For the comparison, we used thermal conductivity of $1.3$~${\rm W/(m\cdot K)}$ which corresponds to a medium-density concrete ~\cite{ISO10456}. Both simulation and analytical calculation give thermal transmittance of $0.15$~${\rm W/(m^2\cdot K)}$ validating our model.

\begin{table}[t]
	\begin{small}
	\begin{center}
		\caption{Thermal conductivities of materials used in embedded antenna system}
		\label{table:therm_param}
		\begin{tabular}{c|c} \hline
			 Material  &  Thermal conductivity at $20^\circ$C\\ & $\lambda~{\rm [W/(m\cdot K)]}$\\ \hline
			Rogers RT/duroid 5880  &  0.2\\ \hline
			Styrofoam (EPS)  &  $5 \times 10^{-2}$\\ \hline
			Stainless steel  &  15\\ \hline 
			Copper  &  400\\ \hline 
			Teflon (PTFE)  &  0.24\\ \hline 
			Rockwool & $3.5 \times 10^{-2}$
		\end{tabular}
	\end{center}
	\end{small}
\end{table}

\section{Antenna system}
\label{sec:antenna_system}

\subsection{Design of a spiral antenna system}
\label{sec:antenna}

	\begin{figure}[ht]
		\begin{center}
			\includegraphics[scale=0.6]{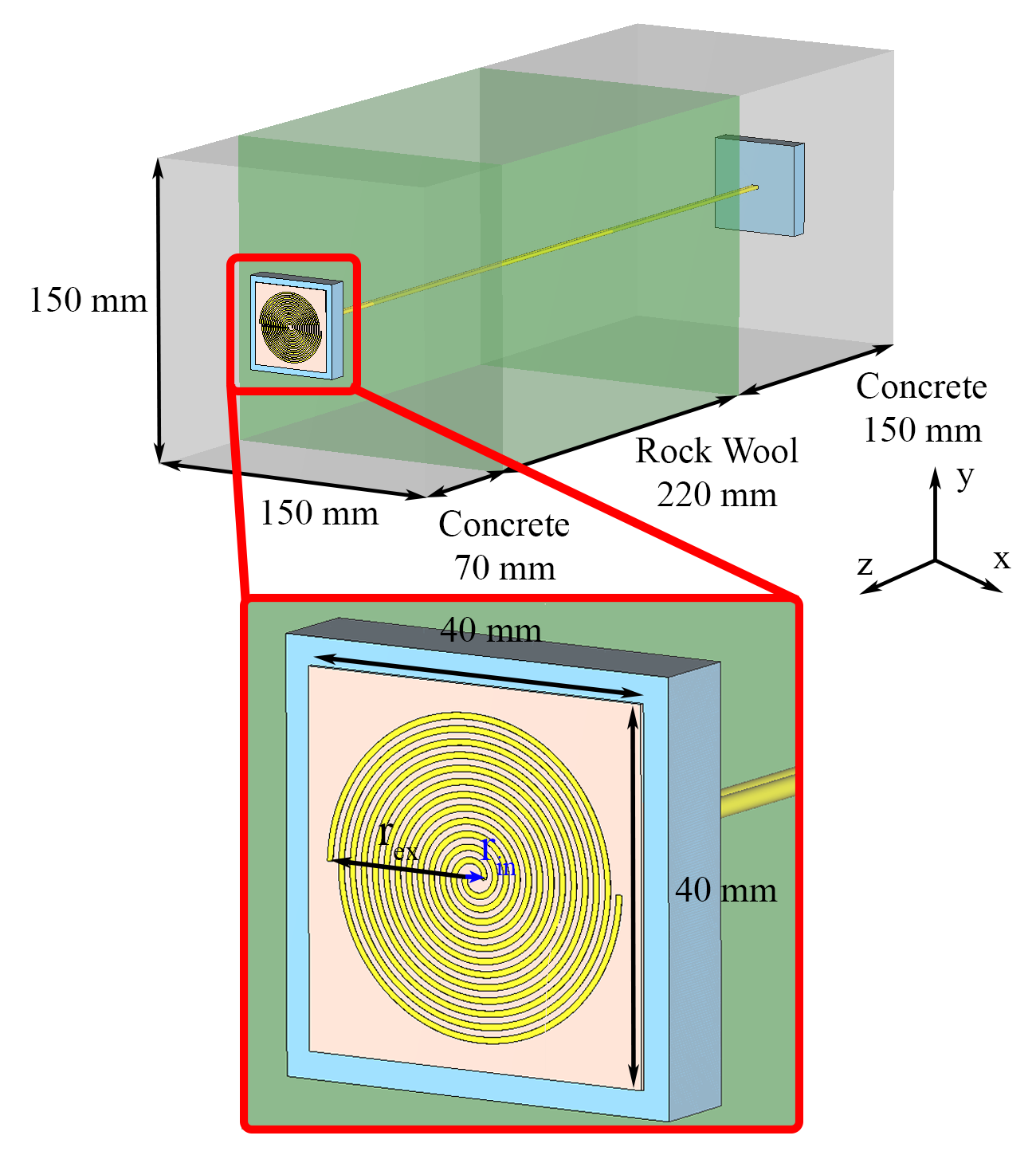}
			\caption{Unit cell of load bearing wall with embedded back-to-back spiral antenna system.}\label{fig:system}
		\end{center}
	\end{figure}
	
Our antenna system contains two antenna elements located on different sides of the wall connected back-to-back with coaxial cables, similar to our earlier work~\cite{vahasavo_EuMC21}. However, the earlier work uses patch antennas that cover only $1.3$~\% of the relative bandwidth. Wideband antennas are preferable over narrow ones since those can cover a wide variety of cellular services. The antenna system embedded into the load bearing wall is shown in Fig.~\ref{fig:system} where the Archimedean spiral antennas are designed and installed instead of the patch antennas. We chose the spiral antenna because of its compact size. Compared to e.g. bowtie antenna, the size of the spiral antenna is roughly half which makes it possible to embed them in the wall densely. As the spiral antennas are electrically balanced, a dual-coaxial cable is introduced to connect two antennas on different sides of the wall. The outer conductors of two coaxial cables are galvanically connected, making them a balanced line with double the characteristic impedance of the original coaxial cable. This way a similar characteristic impedance is realized for the spiral antennas and the dual-coaxial cable, thereby omitting the need for a matching network or balun. Each coaxial cable has a center pin with $0.287$~mm diameter and an outer connector with $1.76$~mm diameter and $0.2$~mm thickness. Low-density Teflon (PTFE) with a dielectric constant of $1.75$ and ${\rm tan}\delta = 0.004$ is used as an insulating layer of cables. A single and dual-coaxial cable, therefore, shows 82~$\Omega$ and 164~$\Omega$ characteristic impedance. As reported in~\cite{vahasavo_EuMC21}, the thermal transmittance of the wall is too high if copper coaxial cables are going through the whole wall. The thermal conductivity of stainless steel is about $27$ times smaller than copper, as shown in Table~\ref{table:therm_param}. This makes it a suitable conductor material for coaxial cables going through the wall. According to CST simulations, a coaxial cable loss is $3.7$~dB and $6.3$~dB at $3.5$ and $8$~GHz respectively, while the loss of bare load bearing wall is $23.2$~dB and $42.5$~dB at $3.5$ and $8$~GHz, showing much greater losses of wave propagation through the wall compared to the coax cable of the same length. Because the thermal conductivity of stainless is low, the cables cannot be soldered together so we decided to place them inside heat shrink tubing. The 3D model of the unit cell of the antenna system embedded into the wall can be downloaded from~\cite{vaha_savo_2023_zenodo}

\subsection{Spiral antenna element and its measurements in free space}
The spiral antenna is realized on top of $0.5$~mm thick Rogers RT/duroid 5880 laminate ($\epsilon_{\rm r}=2.2$, $\tan\delta= 9.0 \times 10^{-4}$) with the size of $40\times 40$~mm. The legs of the spiral are realized with lines of $0.68$~mm width. The number of turns in the spiral is 6 with the internal and external radius of $r_{\rm in}=1.08$~mm and $r_{\rm ex}=17.4$~mm. Since the internal radius is the same as the radius of our coaxial cable, each leg of the spiral antenna can be connected to the center pins of the dual-coaxial cable to build the back-to-back antenna systems. 

\begin{figure}[ht]
		\begin{center}
			\includegraphics[scale=0.35]{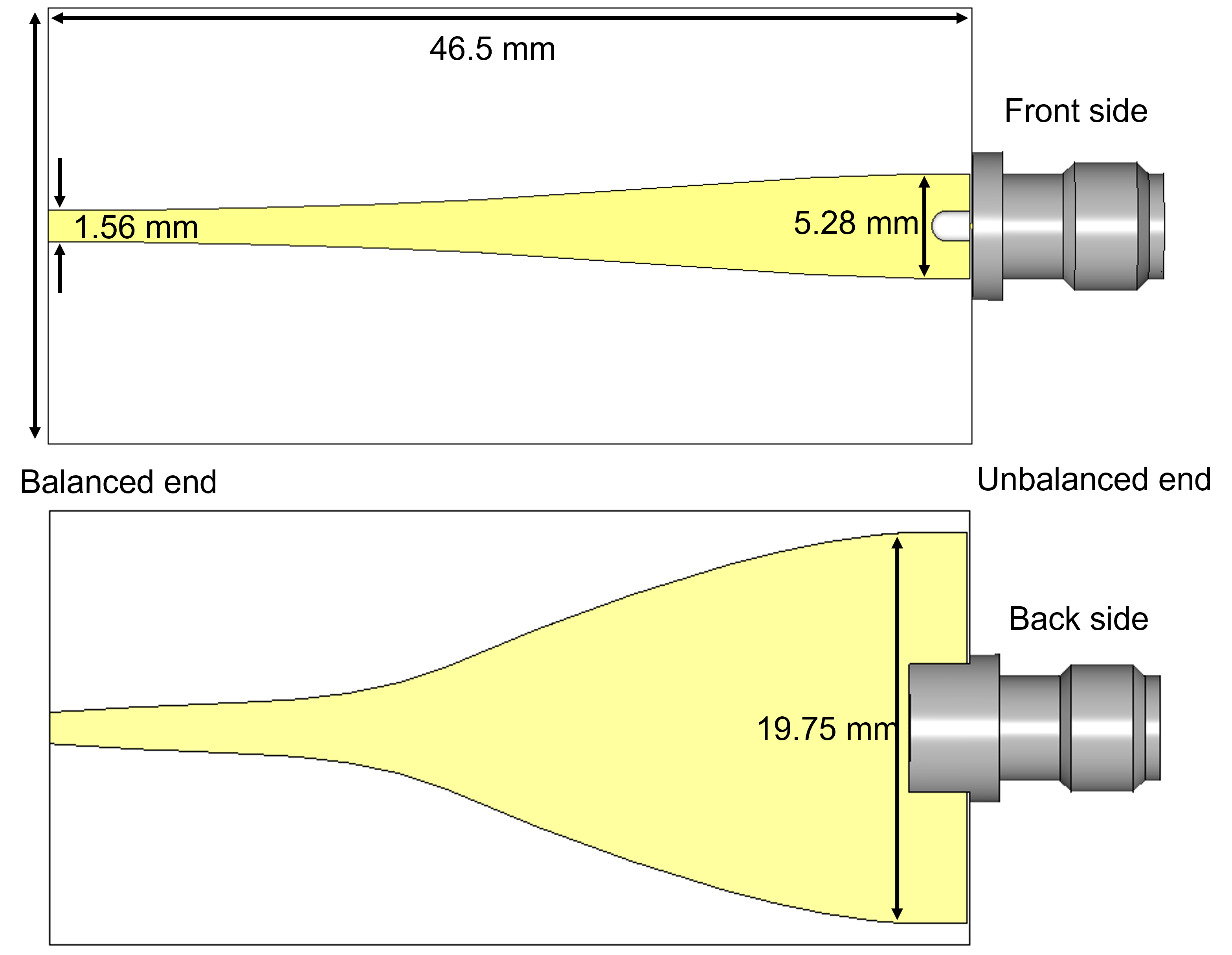}
			\caption{Front and back view of experimentally tapered microstrip balun. Supporting legs on the top side of the connector are removed to ensure that the connector does not touch the feed line of the balun.}
			\label{fig:balun}
		\end{center}
	\end{figure}

To be able to test the spiral antenna in free space, without the influence of a wall, a balancing unit (balun) is needed. An experimentally tapered balun, similar to the one in \cite{Vinayagamoorthy_VTC12}, was designed. The balun is realized on top of $1.575$~mm thick Rogers RT/duroid 5880 laminate with the size of $22\times 46.5$~mm. The unbalanced end of the balun is a transmission line with $5.28$~mm width to have $50$~$\Omega$ input impedance. The tapering of the line and the ground plane leads to a parallel plate line of $1.595$~mm width so that the input impedance at the balanced end is $164$~$\Omega$ which is similar to the input impedance of the spiral antenna and the dual-coaxial cable. The ground plane printed on the backside of the balun is $19.75$~mm wide on the unbalanced end and is tapered to the same width as the feed line. The top and bottom sides of the balun are shown in Fig.~\ref{fig:balun}.

\begin{figure}[ht]
		\begin{center}
				{\includegraphics[scale=0.55]{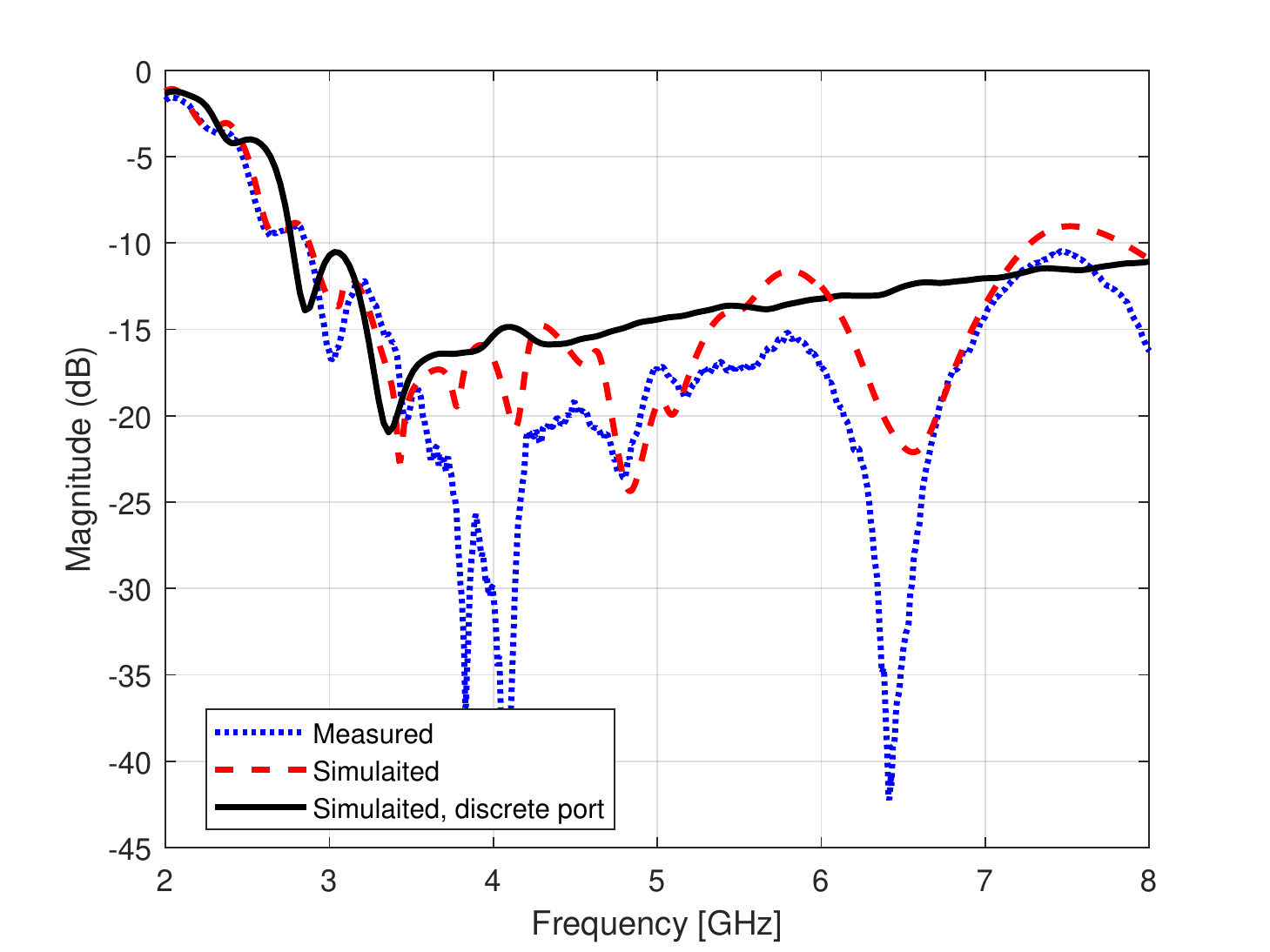}}
			\caption{Reflection coefficient of the spiral antenna in free space, measured with the balun.}\label{fig:spiral_balun_S}
		\end{center}
	\end{figure}

Two vias were used to connect each leg of the spiral antenna and the balun. Via dimensions are similar to the center connectors of the dual-coaxial cable to ensure a similar current transition with balun and dual-coaxial cable. Amphenol SV Microwave $2.92$~mm connector (Mfr. No: 1521-60051) was used to connect the antenna to the VNA. 
To make sure that the connector does not touch the feed line, supporting legs on the top side of the connector are removed, as can be seen from Fig.~\ref{fig:balun}.  Figure \ref{fig:spiral_balun_S} shows the measured and simulated reflection coefficient of the balun attached to the spiral antenna between $2$ and $8$~GHz covering most of the NR FR1. The analysis is limited below $8$~GHz since we do not have a powerful computational capability enough to simulate the infinitely large antenna embedded wall above $8$~GHz. The simulated reflection coefficient is smaller than the measured one over the whole frequency band. This could be caused by the manufacturing tolerances since the balun is experimentally tapered and small changes in balun dimensions can improve the balun. Both simulated and measured curves show ripples over the frequency band. This is most likely caused by imperfect connector and soldering which can cause small changes to the matching level over the frequency range. Also, the matching between the antenna and balun can vary a little over the frequency. The black line corresponds to the simulation with a discrete port with $164$~$\Omega$ normalization impedance, i.e., that of the dual-coaxial cable. All simulated and measured reflection coefficients agree well with each other.

\begin{figure}[ht]
		\begin{center}
					{\includegraphics[scale=0.55]{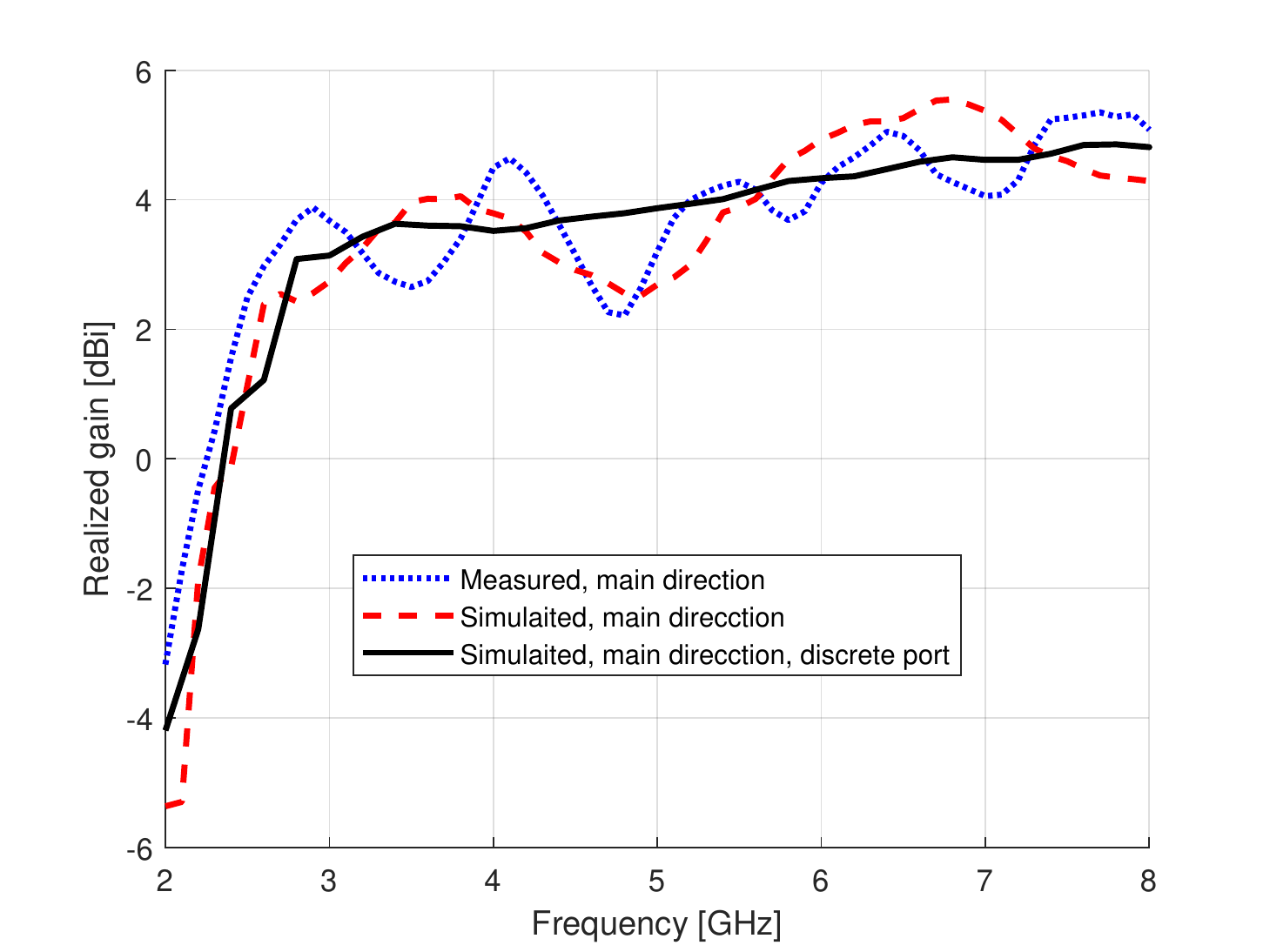}}
			\caption{The realized gain of the spiral antenna with the balun in free space.}\label{fig:free_gain}
		\end{center}
	\end{figure}

The realized gain of the spiral antenna was measured in an anechoic chamber between $2$ and $8$~GHz. Figure \ref{fig:free_gain} shows the maximum of realized gain of the spiral antenna over the frequency to broadside direction $+z$. The simulation with a discrete port has a much smoother gain pattern over the frequency, suggesting that the balun causes some losses and standing waves to the system. The standing waves can be caused by e.g. imperfect matching between the antenna, balun, and VNA. Although the measured maximum gain fluctuates more than the simulated one, the overall trend of the realized gains agrees with each other, confirming the efficacy of the designed spiral antenna.

\section{Antenna embedded wall}

\subsection{Embedding antenna systems on the wall}

The spiral antenna cannot be placed directly on top of concrete because concrete causes detuning and losses to the antenna. At the same time, in free space, a spiral antenna radiates the same way in $\pm z$ directions on Fig.~\ref{fig:system}. To improve the radiation efficiency of the antenna and direct more power to the $+z$-direction defined in Fig.~\ref{fig:system}, the antenna is backed with the Rohacell foam when integrated with a wall. The foam is placed inside of the concrete layer to create a small air gap between the antenna and the concrete hence maintaining the high radiation efficiency of the antenna over the frequency. The thickness of the foam was optimized in simulations to maximize the radiation efficiency when the spiral antenna was placed on top of a $150\times 150\times 50$~${\rm mm^3}$ concrete slab. The antenna element was backed with a block of foam with the size of $50\times 50$~${\rm mm^2}$. The thickness was varied from 0 to 10~mm. The antenna was fed by the dual-coaxial cable described in Section~\ref{sec:antenna}. The radiation efficiency increases monotonically when the foam layer is thicker. Without the foam layer, the average radiation efficiency of the antenna between $2.5$ and $8$~GHz is $-12$~dB and a mean realized gain of $-6.4$~dBi towards $+z$-direction. When we add $10$~mm thick foam to the backside of the antenna the average radiation efficiency is $-3.1$~dB and the mean realized gain of $4.6$~dBi, thereby choosing $10$~mm thick Rohacell in our back-to-back antenna system. Because of the limited capability to feed differential signals with our measurement equipment, we cannot verify these results with measurements. The operation of the antenna element on top of the concrete is tested by measuring the antenna system embedded into a wall.

\subsection{Electromagnetic transmission of the infinitely large antenna-embedded wall}
\label{sec:Electromagnetic_transmission}

	\begin{figure}[ht]
		\begin{center}
				{\includegraphics[scale=0.55]{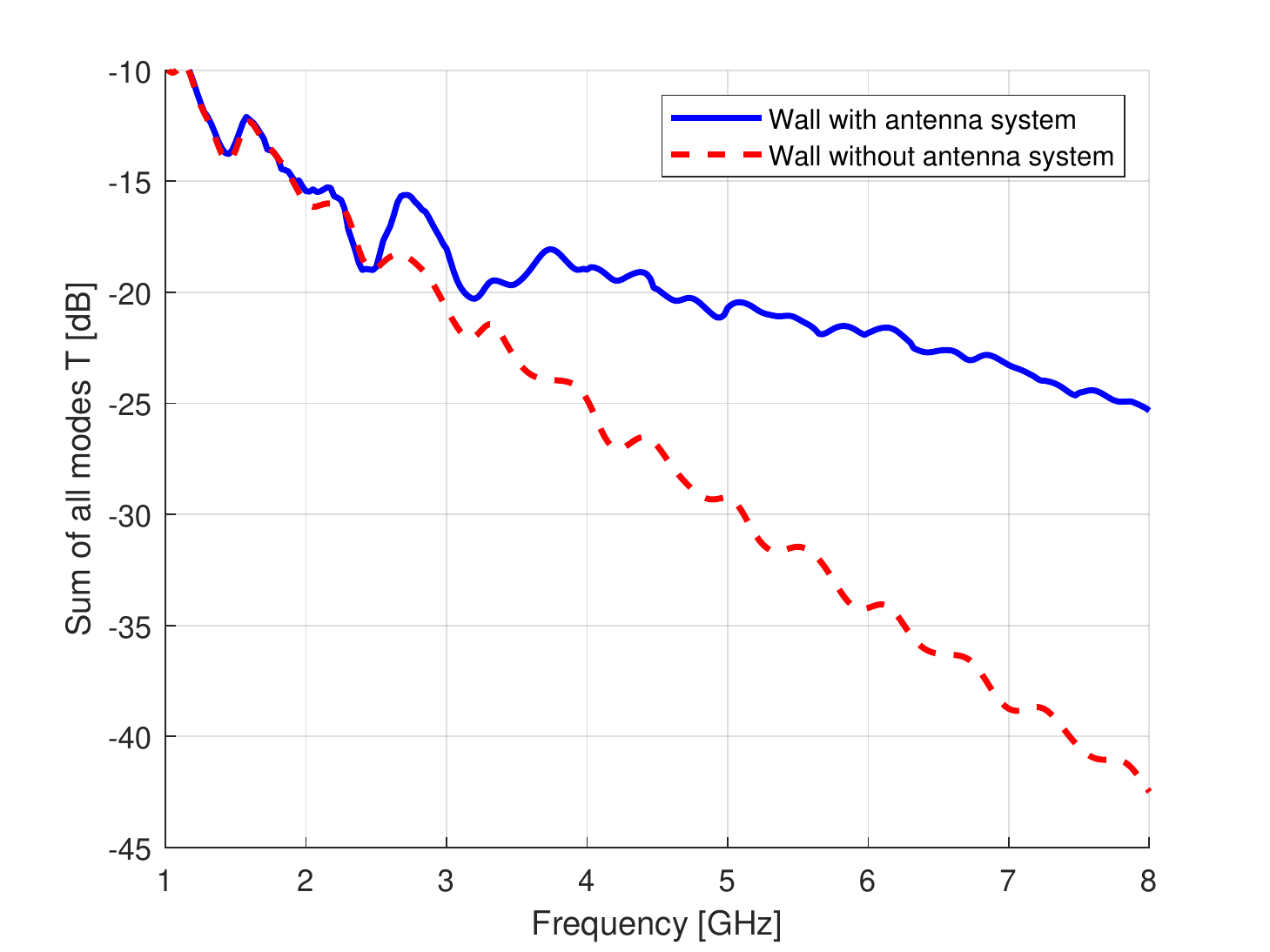}}
			\caption{Simulated transmission through the infinitely large load bearing wall with and without the embedded spiral antenna system. The unit cell size is in $x$ and $y$ directions $150$~mm. }\label{fig:transmission_unit}
		\end{center}
	\end{figure}

\begin{figure}[ht]
	\begin{center}
		\subfigure[]{\includegraphics[scale=0.28]{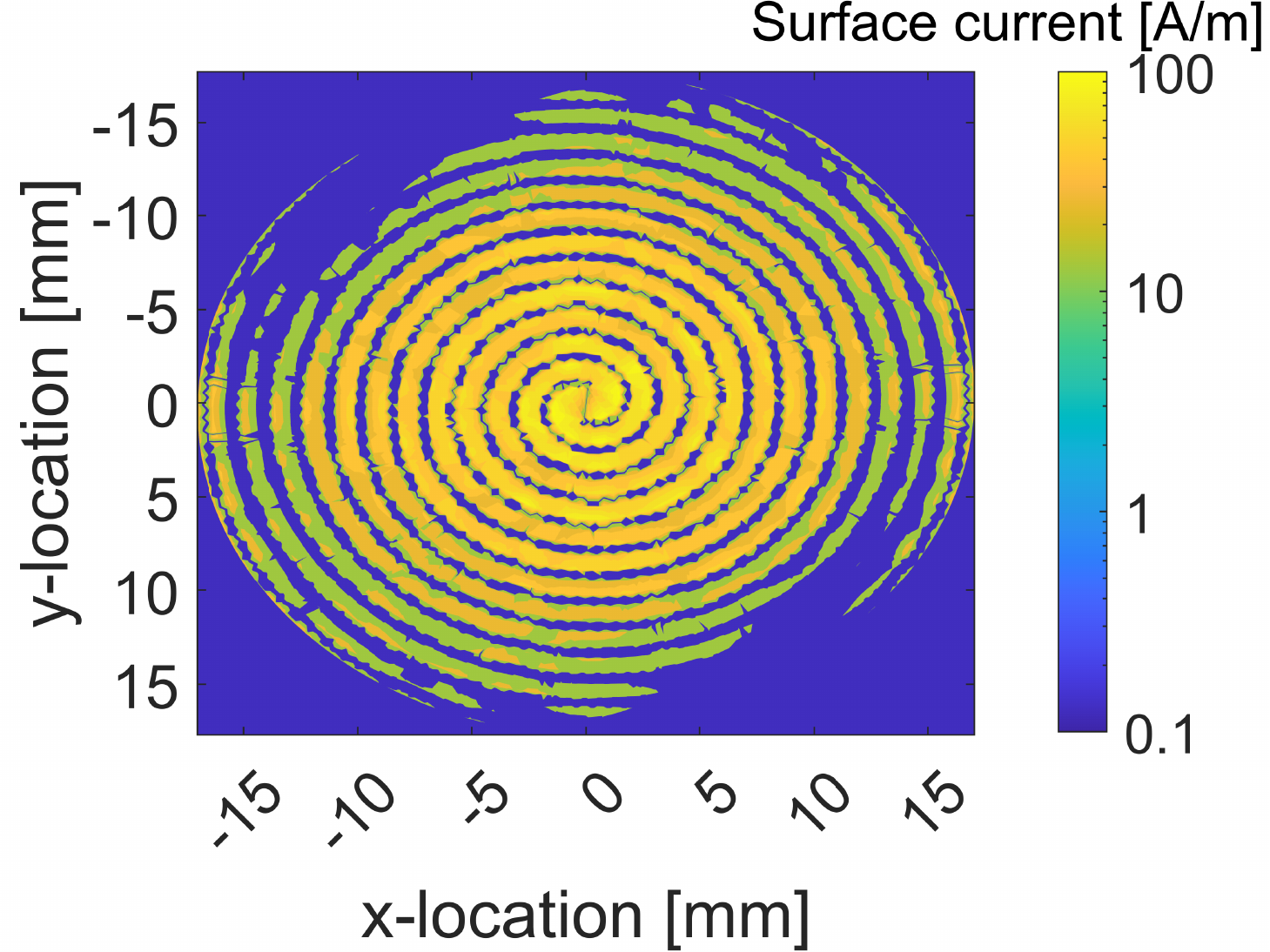}
			}
		\subfigure[]{\includegraphics[scale=0.28]{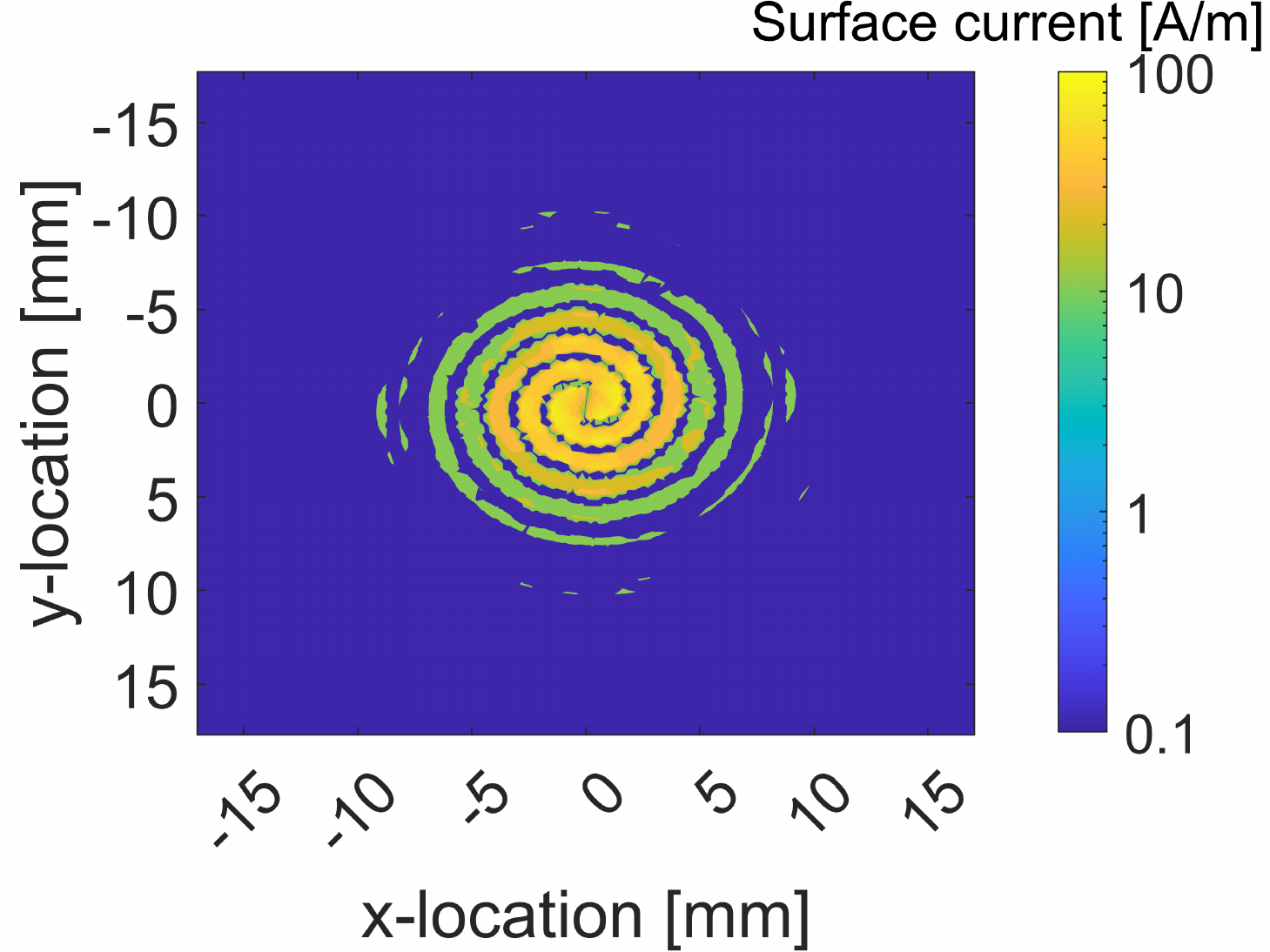}
			}
		\caption{Surface current of the spiral antenna at (a) $3$~GHz, (b) $8$~GHz.
		}
		\label{fig:sufaceCurrent}
	\end{center}
\end{figure}

The total electromagnetic transmission of the wall was simulated with and without an antenna system as explained in Section~\ref{sec:Electromagnetic_insulation}. Figure \ref{fig:transmission_unit} shows the transmission coefficient from $1$ to $8$~GHz. The unit cell size is $150\times 150$~${\rm mm^2}$. It is important to remember that the unit cell dimensions in $x$ and $y$ directions are the same as the antenna separations on the infinitely large wall. We used a right-handed circular polarized incident plane wave. As can be seen from the Figure, the antenna system starts to improve the transmission around $2.6$~GHz and above. The transmission coefficients decrease as the frequency increases even after embedding the antenna systems. This is due to the increasing cable loss and also decreasing effective aperture size of the spiral antenna. A decrease in the effective aperture can be observed from the simulated surface current plot at $3$ and $8$~GHz depicted in Fig.~\ref{fig:sufaceCurrent} At $8$~GHz currents flow in roughly $5$ times smaller area than at $3$~GHz. Up to $17$~dB improvement of the transmission coefficient was observed at $8$~GHz by embedding the antenna systems to the wall. Compared to our previous work, the bandwidth of the antenna system is improved from 1.3~\% to 3.1:1. At the same time, both antenna systems realize a similar improvement of the transmission coefficient at $3.5$~GHz. 

\begin{figure}
    \centering
    {\includegraphics[scale=0.55]{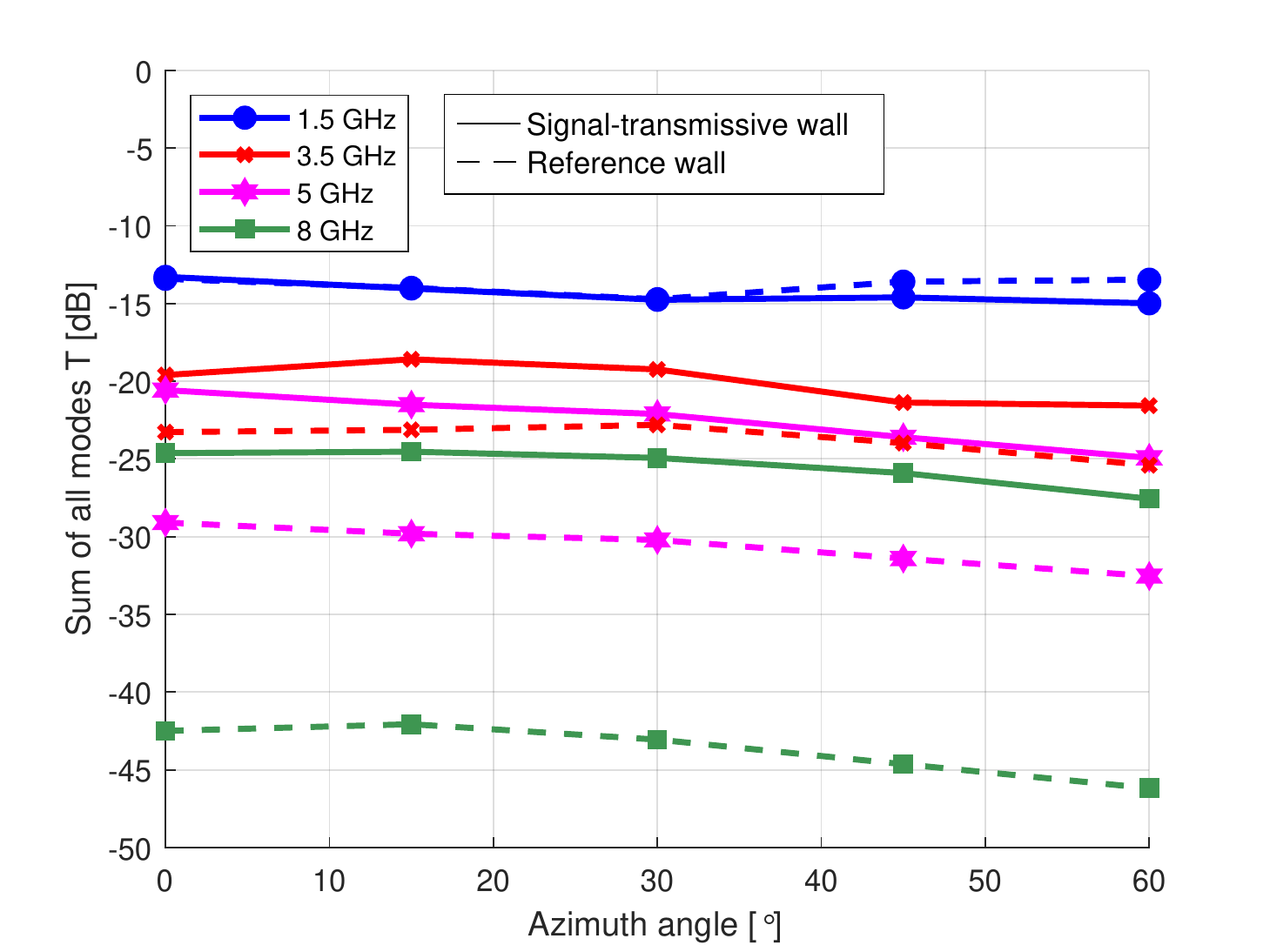}}
    \caption{Simulated transmission through the infinitely large load bearing wall with and without the embedded spiral antenna system with a changing incident azimuth angle of a plane wave from $0$ to $60^\circ$. T°. The unit cell size spans $150$~mm in $x$ and $y$ directions.}
    \label{fig:incident}
\end{figure}

Next, the effect of azimuth incident angle is studied at $1.5,\  3.5,\  5$, and $8$~GHz. For this study, we need to increase the number of simulated Floquet modes because of the increased incident angle. Also, scattering caused by antennas and cables will increase the number of propagating modes through the wall. Figure~\ref{fig:incident} shows the transmission coefficient in the case of $15^\circ,\ 30^\circ,\ 45^\circ$ and $60^\circ$ incident angles. The figure shows that transmission through the wall with and without the antenna system is decreasing when the azimuth incident angle is increased. This is expected since we know from~\cite{ITU-R_P2040} that the reflection coefficient of concrete is increasing when the incident angle is greater. Also, the gain of the spiral antenna is decreasing with greater incident angles. The antenna system clearly improves the transmission through the wall even with the glazing incident angle. For example, at $8$~GHz with $60\circ$ incident angle, the improvement is $18.6$~dB which is $1.6$~dB more than in the case of a normal incident. Measuring these results would demand us to manufacture a large signal-transmissive wall which is not possible in our laboratory. Instead of using multiple incident angles in the measurements, we concentrate on showing the signal-transmissive wall's efficacy in the case of a normal incident.

\subsection{Manufactured wall samples}

	\begin{figure}[ht]
		\begin{center}
			\includegraphics[scale=0.19]{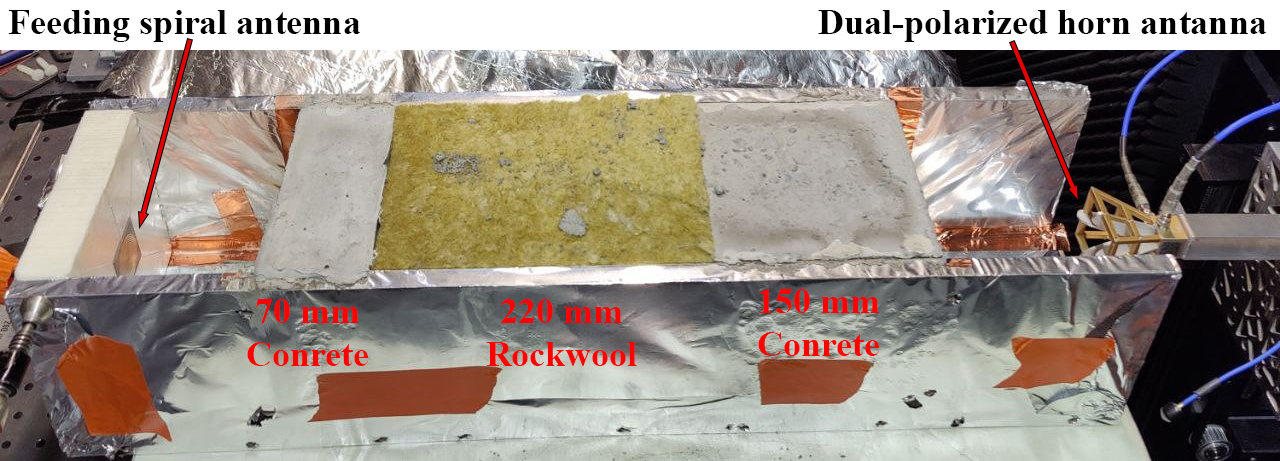}
			\caption{The measurement setup for the electromagnetic transmission measurements. The left side is the feeding spiral antenna and on the right side, the dual-polarized horn antenna. The top-side metal is removed for the picture.}
			\label{fig:EMmesSetup}
		\end{center}
	\end{figure}

As it is not possible to manufacture infinitely large wall as was performed in the simulations, wall samples of $150\times 150~{\rm mm^2}$ unit cell size were manufactured for our measurements. These small samples are enclosed in metallic waveguides for electromagnetic measurements. The waveguides are used {\it only} for the purpose of comparing the simulated and measured antenna-embedded walls to verify its whole design. If the measurements agree with simulations when the antenna-embedded wall is enclosed by the metallic waveguide, the same should apply to other setups of the antenna-embedded walls. The wall samples are cast inside the waveguides to ensure that the samples are as close to the sizes of the waveguides as possible. The waveguide-enclosed antenna embedded wall being the device-under-test, the same spiral antenna with the balun that we used in free space measurements was used as a feed antenna. As a receiving antenna, we use a dual-polarized double-ridge horn antenna. The setup using the waveguide allows for measuring all the energy that is transmitted through the wall sample. The same setup was built in a simulation for comparison with measurements, where a waveguide port with a large enough amount of modes is used as a receiving antenna to ensure that all the transmitted power is detected. 

The load bearing wall consists of a rockwool layer in addition to concrete. To ensure the correct thermal behavior of the wall, the Paroc COS 5ggt rockwool was used, which is widely used in concrete sandwich elements in Finland. The antenna system and rockwool were placed inside the waveguide and the concrete was cast in. Also, a wall without the antenna system was cast as a reference sample.  The waveguide has  $150$~mm extra length at both ends of the wall sample, leading to the total length of the waveguides being $740$~mm. The empty space created by the extra length of the waveguide is needed for placing the transmitting and receiving antennas. The curing time of concrete varies based on the temperature and humidity of the casting site. Concrete reaches its final strength roughly in 28 days \cite{ACI_318-19} so any measurements were performed after that period.

	\begin{figure}[ht]
		\begin{center}
				{\includegraphics[scale=0.55]{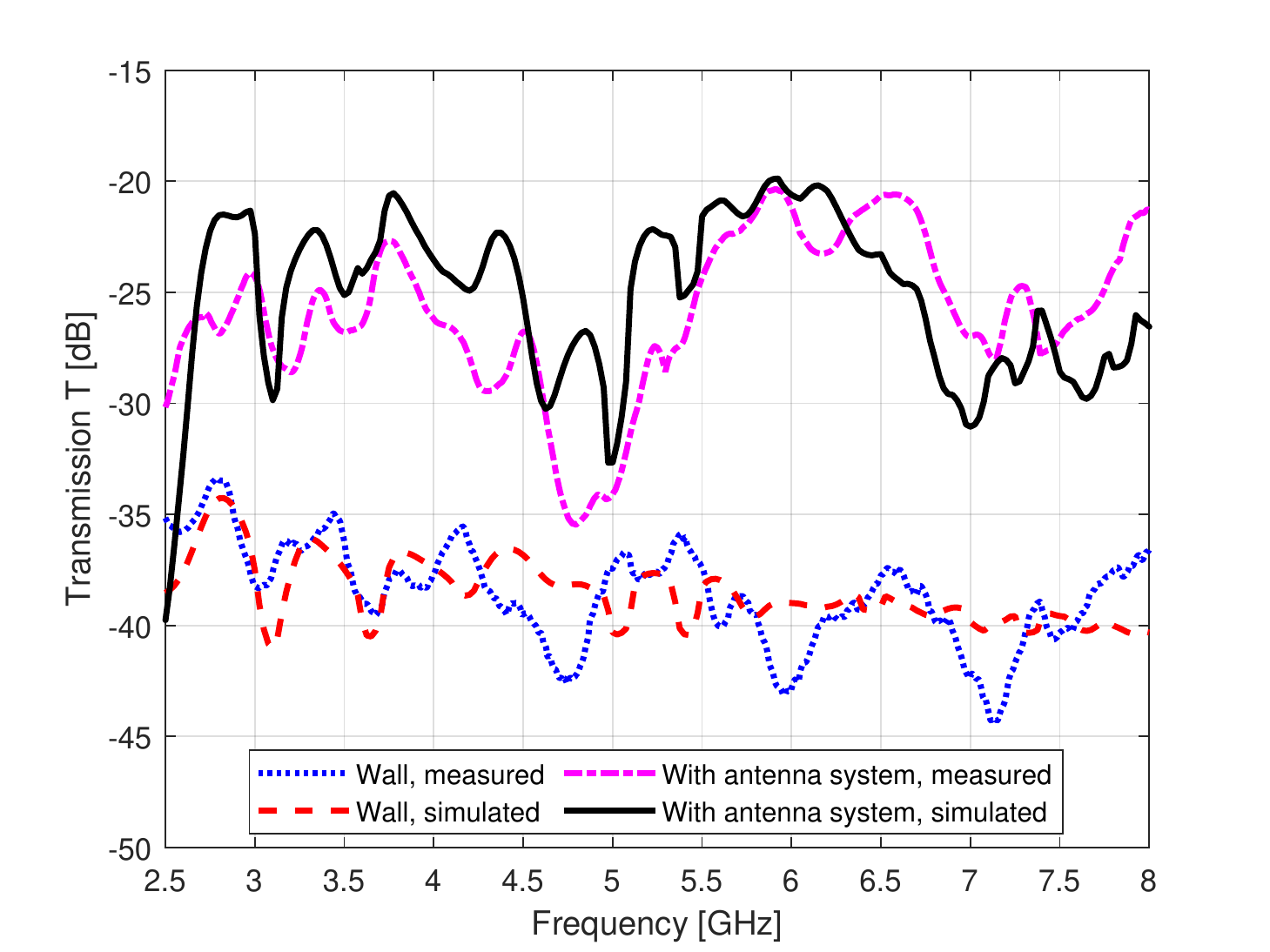}}
			\caption{Measured and simulated total electromagnetic transmission through the wall sample inside the waveguide.}\label{fig:transResults}
		\end{center}
	\end{figure}

\subsection{Electromagnetic insulation of the antenna-embedded wall inside a waveguide}

Prior to electromagnetic measurements of the antenna-embedded wall, a transmission coefficient of {\it only} the casted concrete block in the waveguide was measured to estimate the complex permittivity of the concrete. It turned out that the casted concrete had a moisture content that led to higher permittivity than the one in the ITU model. The permittivity was estimated by solving the analytical model of transmission coefficients for a slab shown in ITU-R p.2040~\cite{ITU-R_P2040}. The estimated parameters for concrete to be used in equation~\ref{eq:e} are $a=5.84$, $b=0$, $c=0.205$, $d=0.06$. We use this measured permittivity to compare the transmission through the reference and antenna-embedded walls between the measurements and simulations.

Figure \ref{fig:EMmesSetup} shows the measurement setup used in the electromagnetic transmission measurements without the top-side waveguide wall. The feeding spiral antenna is mounted at the center of a piece of Styrofoam. The corners of the waveguide are strengthened with copper tape. During the measurements, the top side of the waveguide wall is also used. Figure~\ref{fig:transResults} shows the total transmission through the wall with and without the spiral antenna system. The losses caused by the measurement system, including the waveguide, are calibrated by normalizing the wall measurement coefficients with those with an empty waveguide of the same dimension. The measured and simulated transmission agrees well with each other. The improvement due to the inclusion of antenna systems is on average $16$~dB between $2.5$ and $8$~GHz. The two dips in the transmission around $4.8$ and $7.3$~GHz are caused by feed antenna coupling to the embedded antenna system. These dips would be smaller if the distance between the feed antenna and the embedded antenna would be larger. The impact of the antenna system in improving electromagnetic transmission through the load bearing wall is clear.

\subsection{Thermal insulation of the antenna-embedded wall}
\label{sec:thermal}

The thermal conductivities of most materials used in the signal-transmissive wall are given by manufacturers as listed in Table~\ref{table:therm_param}. But the concrete was mixed in our lab and its thermal properties are unknown. 
ISO10456~\cite{ISO10456} gives nominal thermal properties of most used construction materials. A sample of our concrete was measured with \textit{C-Therm TCI thermal conductivity measurement system}\footnote{https://ctherm.com/thermal-conductivity-instruments/tci-thermal-conductivity-analyzer} to characterize the thermal conductivity of our concrete. C-Therm TCI is based on a modified transient plane source technique, requiring only one probe. A single measurement of thermal conductivity takes only a few seconds. The measurement was repeated a few times to make sure that the results are repeatable. The thermal conductivity estimate of our concrete was $1.3$~${\rm W/(m\cdot K)}$ which is similar to medium density concrete in~\cite{ISO10456}. The measured thermal conductivity of concrete is used for the following thermal simulations.

The thermal insulation of the wall was numerically simulated with the same structures as the electromagnetic unit cell simulations using the method described in Section~\ref{sec:Electromagnetic_insulation}. The wall without the antenna system has a U-value of $0.15$~${\rm W/(m^2\cdot K)}$, while that of the antenna embedded wall is $0.16$~${\rm W/(m^2\cdot K)}$, showing a slight increase. The present U-value of the antenna embedded wall is better than the one in our previous work~\cite{vahasavo_EuMC21}, where we reported a U-value of $0.51$~${\rm W/(m^2\cdot K)}$, most likely because the stainless steel conductor was used in coaxial cables instead of copper. Also, the reduced size of the cables decreases the U-value. The presented antenna embedded wall has a lower U-value than $0.17$~${\rm W/(m^2\cdot K)}$ limit given in the national building regulation~\cite{buildingcode}.

\subsection{Impacts of Antenna System Separation on Electromagnetic and
Thermal Insulation}

\label{sec:separation}
	\begin{figure}[ht]
		\begin{center}
			{\includegraphics[scale=0.55]{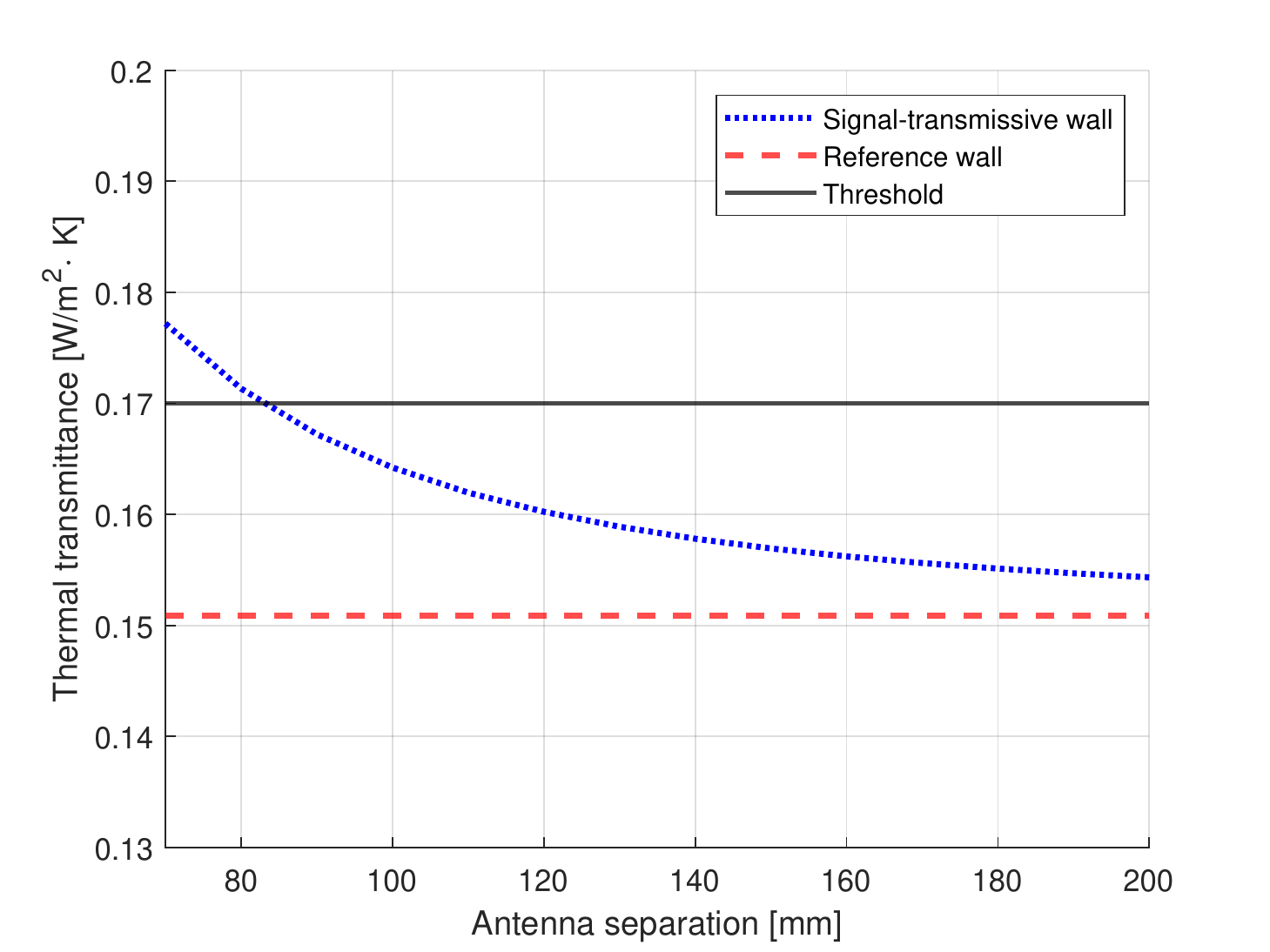}}
			\caption{Simulated thermal transmittance through the infinitely large load bearing wall with and without the embedded spiral antenna system. The antenna separation varies from $70$ to $200$~mm. }\label{fig:u-value_unitcell}
		\end{center}
	\end{figure}

Now that the analysis methods of the signal-transmissive wall are verified through experiments, we are ready to study the effect of antenna separation on electromagnetic transmission and thermal transmittance. When antenna systems are placed more densely on the wall the U-value will increase. The national building code of Finland says that the U-value of the wall cannot be higher than $0.17$~${\rm W/(m^2\cdot K)}$~\cite{buildingcode}, as mentioned in the previous section. The antenna separation is first studied in terms of thermal transmittance by changing the squared unit cell size in $x$ and $y$ directions by varying their values from $70$ to $200$~mm  to identify the smallest possible cell size that meets the $0.17$~${\rm W/(m^2\cdot K)}$ limit. Figure~\ref{fig:u-value_unitcell} plots the results, showing that $90$~mm antenna system separation satisfies the $0.17$~${\rm W/(m^2\cdot K)}$ threshold.

	\begin{figure}[ht]
		\begin{center}
			{\includegraphics[scale=0.55]{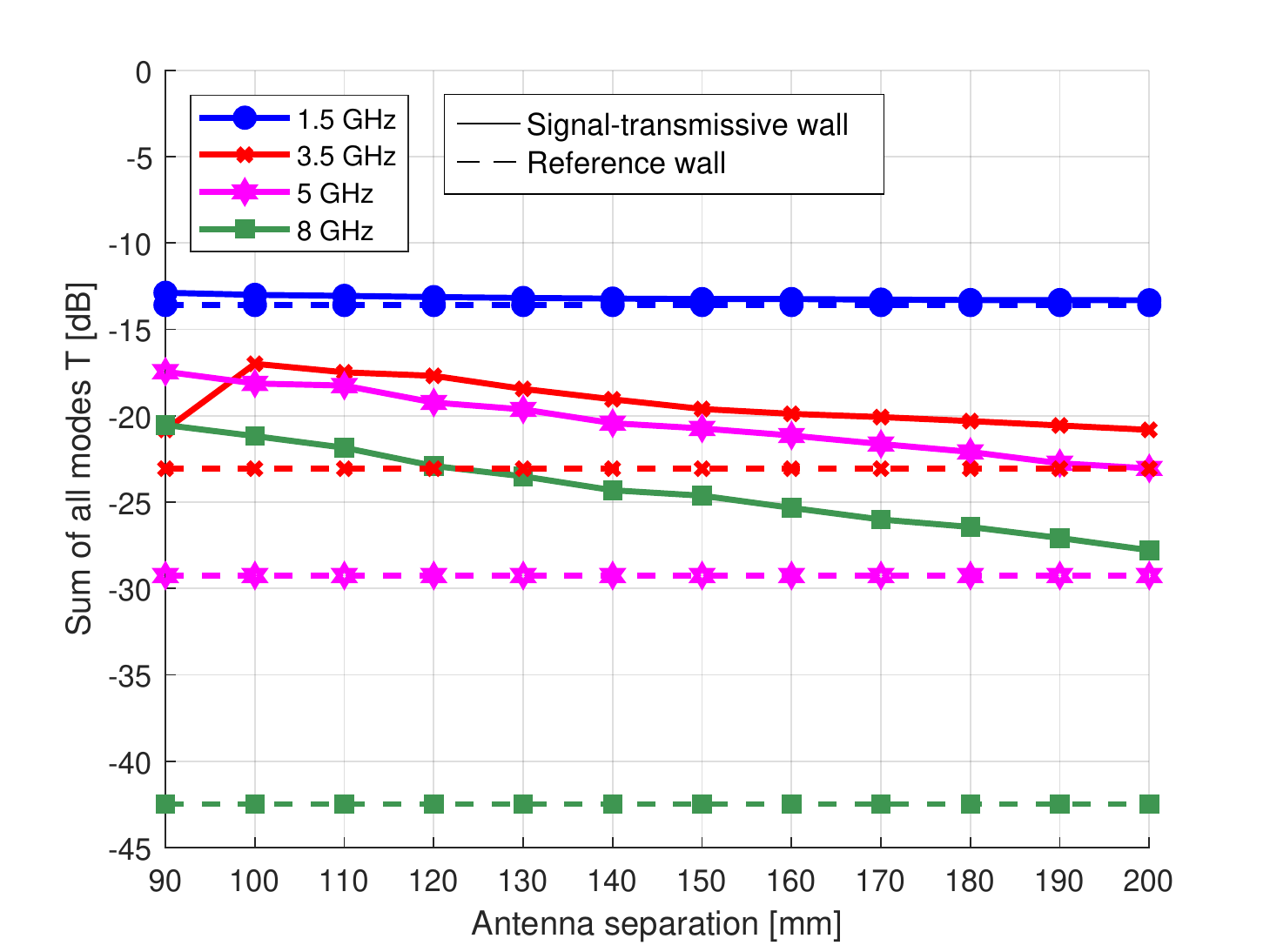}}
			\caption{Simulated transmission through the infinitely large load bearing wall with and without the embedded spiral antenna system; the antenna system separation varies from $70$ to $200$~mm. }\label{fig:antennaseparation}
		\end{center}
	\end{figure}

Next, the electromagnetic transmission through the wall is simulated for varying unit cell dimensions from $90$ to $200$~mm, similar to thermal simulations. The RF was $1.5$, $3.5$, $5$ and $8$~GHz. Figure~\ref{fig:antennaseparation} shows the results including those of raw load bearing wall. The transmission coefficient is higher with smaller antenna system separations except at $3.5$~GHz where $90$~mm antenna separation gives $4$~dB lower transmission than $100$~mm separation. This can be explained by the fact that antenna elements are coupled to each other with small separations. For a constant antenna separation, greater effective aperture size leads to stronger coupling as evidenced in Fig.~\ref{fig:sufaceCurrent}. At $8$~GHz the transmission is improved by $22$~dB  with $90$~mm antenna separation which is roughly $5$~dB more than with the $150$~mm separation that was presented in Section~\ref{sec:Electromagnetic_transmission}. According to these results, $100$~mm antenna system separation gives us the lowest transmission loss through the wall with the current design of the signal-transmissive wall. With $100$~mm separation the mutual coupling does not deteriorate the radiation efficiency of the antenna system while the signal-transmissive wall respects the U-value restrictions. It is also good to notice that even with $200$~mm antenna system separation the transmission coefficient is improved at all the simulated RF.

\section{Concluding Remarks}
\label{sec:conclusion}

This paper introduces analytical, numerical, and empirical evaluation of the antenna-embedded wall, which is called the signal-transmissive wall. Wideband spiral antenna system was introduced, manufactured and its electromagnetic and thermal transmission characteristics were analyzed. Dual-coaxial cable assembly was introduced to connect two balanced spiral antennas back-to-back without the need for a separate balun or matching network. Penetration loss of load bearing wall was decreased down to $17$~dB at $8$~GHz when the antenna system is embedded to the wall every $150$~mm, compared to $42.5$~dB loss of the bare load bearing wall. The electromagnetic transmission is improved by more than $6$~dB at RF above $4$~GHz. The increase of the U-value was so slight that the signal-transmissive wall still achieves the demanded level of national building regulation. The simulated and measured electromagnetic transmissions agree well with each other for the waveguide-enclosed signal-transmissive wall. Antenna systems could be placed as densely as every $100$~mm to the load bearing wall without exceeding the U-value threshold while improving the electromagnetic transmission at $8$~GHz by $22$~dB. Given the verified improvement of the electromagnetic transmission coefficient through the signal-transmissive wall, its impact on indoor cellular coverage inside low-energy buildings is a subject of further study. These studies include an analytical model that allows us to study in more detail the effects of different components, like antennas and cable assemblies, and the materials used on them on the link budget of radio links involving the embedded antenna systems~\cite{Lorenzo_EuCAP23}. We are also doing coverage studies with a ray launcher that allows us to study the effect of the signal-transmissive wall on the outdoor-to-indoor channel.

	\bibliographystyle{IEEEtran}
	\bibliography{ref}
	
\end{document}